\documentstyle[11pt,mrs2003,epsfig]{article}
\graphicspath{ {./FRB-Figs/} } 

\newcommand{\bce}[0]{\begin{center}} \newcommand{\ece}[0]{\end{center}}
\newcommand{\ben}[0]{\begin{enumerate}} \newcommand{\een}[0]{\end{enumerate}}
\newcommand{\bit}[0]{\begin{itemize}} \newcommand{\eit}[0]{\end{itemize}}

\newcommand{\simlt}[0]{{\lower.5ex\hbox{$\; \buildrel < \over \sim \;$}}}
\newcommand{\simgt}[0]{{\lower.5ex\hbox{$\; \buildrel > \over \sim \;$}}}

 \newcommand{\Mpc}{\,{\rm Mpc}}

  \newcommand{\muK}{\mu {\rm K}}

\newcommand{\eg}{{\it e.g. }}

\newcommand{\planck}{{\sc Planck}} \newcommand{\plancks}{{\sc Planck}
}

\newcommand{\hfi}{{\sc HFI}} \newcommand{\hfis}{{\sc HFI} }
\newcommand{\lfi}{{\sc LFI}} \newcommand{\lfis}{{\sc LFI} }

 \newcommand{\maps}{{\sc WMAP} }

\begin{document} 
\title{CMB ANISOTROPIES, COSMOLOGICAL PARAMETERS AND FUNDAMENTAL PHYSICS:
  CURRENT STATUS \& PERSPECTIVES}

\author{Fran{\c{c}}ois R. Bouchet} \affil{Institut d'Astrophysique de Paris, CNRS,
98 bis Bd Arago, Paris, F-75014, France}

\begin{abstract} 
I describe briefly the Cosmic Microwave Background (hereafter CMB) physics
which explains why high accuracy observations of its spatial structure are a
unique observational tool both for the determination of the global
cosmological parameters and to constrain observationally the physics of the
early universe. I also briefly survey the many experiments which have measured
the anisotropies of the CMB and led to crucial advances in observational
Cosmology. The somewhat frantic series of new results has recently culminated
with the outcome of the WMAP satellite which confirmed earlier results, set
new standards of accuracy, and suggested that the Universe may have reionised
earlier than anticipated. Many more CMB experiments are currently taking data
or being planned, with the Planck satellite on the 2007 Horizon poised to
extract all the cosmological information in the temperature anisotropies, and
foray deeply into polarisation.

\end{abstract} 
 
\section{Introduction\label{sec:struct}} 

As we shall see, the analysis of the CMB temperature anisotropies indicated
that the total energy density of the Universe is quite close to the so-called
critical density, $\rho_c$, or equivalently $\Omega = \rho/\rho_c \simeq
1$. We therefore live in a close-to-spatially flat Universe. In agreement with
the indications of other cosmological probes the recent results of the CMB
satellite WMAP \cite{0302207} indicate that about 1/3 of that density appear
to be contributed by matter ($\Omega_M = 0.29 \pm 0.07$), most of which is dark -
i.e. not interacting electromagnetically - and cold - i.e. its primordial
velocity dispersion can be neglected. The usual atoms (the baryons) contribute
less than about 5\% ($\Omega_B \simeq 0.047 \pm 0.006$) \cite{0302209}. If
present, a hot dark matter component does not play a significant role in
determining the global evolution of the Universe.  While many candidate
particles have been proposed for this CDM, it has not yet been detected in
laboratory experiments, although the sensitivities of the latter are now
reaching the range where realistic candidates may lay. The other $\sim$ 60\%
of the critical density is contributed by a smoothly distributed vacuum energy
density or dark energy, whose net effect is repulsive, i.e. it tends to
accelerate the expansion of the Universe. Alternatively, this effect might
arise from the presence in Einstein's equation of the famous cosmological
constant term, $\Lambda$. While this global census, surprising as it may be,
had been around already for some time (see {\S}~\ref{sec:observ}), the WMAP
results have tightened earlier constraints and gave further confidence to the
model. These constraints obtained from the analysis of CMB anisotropies arise
from - and confirm - the current theoretical understanding of the formation of
large scale structures in the Universe, which we now briefly outline.

The spatial distribution of galaxies revealed the existence of these large
scale structures (clusters of size $\sim 5\,\Mpc$, filaments connecting them,
and voids of size $\sim 50\,\Mpc$), whose existence and statistical properties
can be accounted for by the development of primordial fluctuations by
gravitational instability. The current paradigm is that these fluctuations
were generated in the very early Universe, probably during an inflationary
period; that they evolved linearly during a long period, and more recently
reached density contrasts high enough to form bound objects. Given the census
given above, the dominant component that can cluster gravitationally is Cold
Dark Matter (CDM).

The analysis of the CMB anisotropies also indicate that the initial
fluctuations statistics had no large deviations from a Gaussian distribution
and that they where mostly adiabatic, i.e. all components (CDM, baryons,
photons) had the same spatial distribution. The power spectrum of the
``initial conditions'' appears to be closely approximated by a power law $P(k)
=\, <|\delta_k|^2> = A_S\, k^{n_S}$, where $\delta_k$ stands for the Fourier
transform of the density contrast ($P(k)$ is therefore the Fourier transform
of the two-point spatial correlation function). The logarithmic slope, $n_S$,
is quite close to unity ($n_S = 0.99 \pm  0.04$ from WMAP alone
\cite{0302209}). This shape implies that small scales collapsed first,
followed by larger scales, with small objects merging to form bigger
objects. The formation of structures thus appear to proceed {\em
hierarchically} within a ``cosmic web'' of larger structures of contrast
increasing with time.

\begin{figure}[htb] \begin{center} \hbox{
\psfig{file=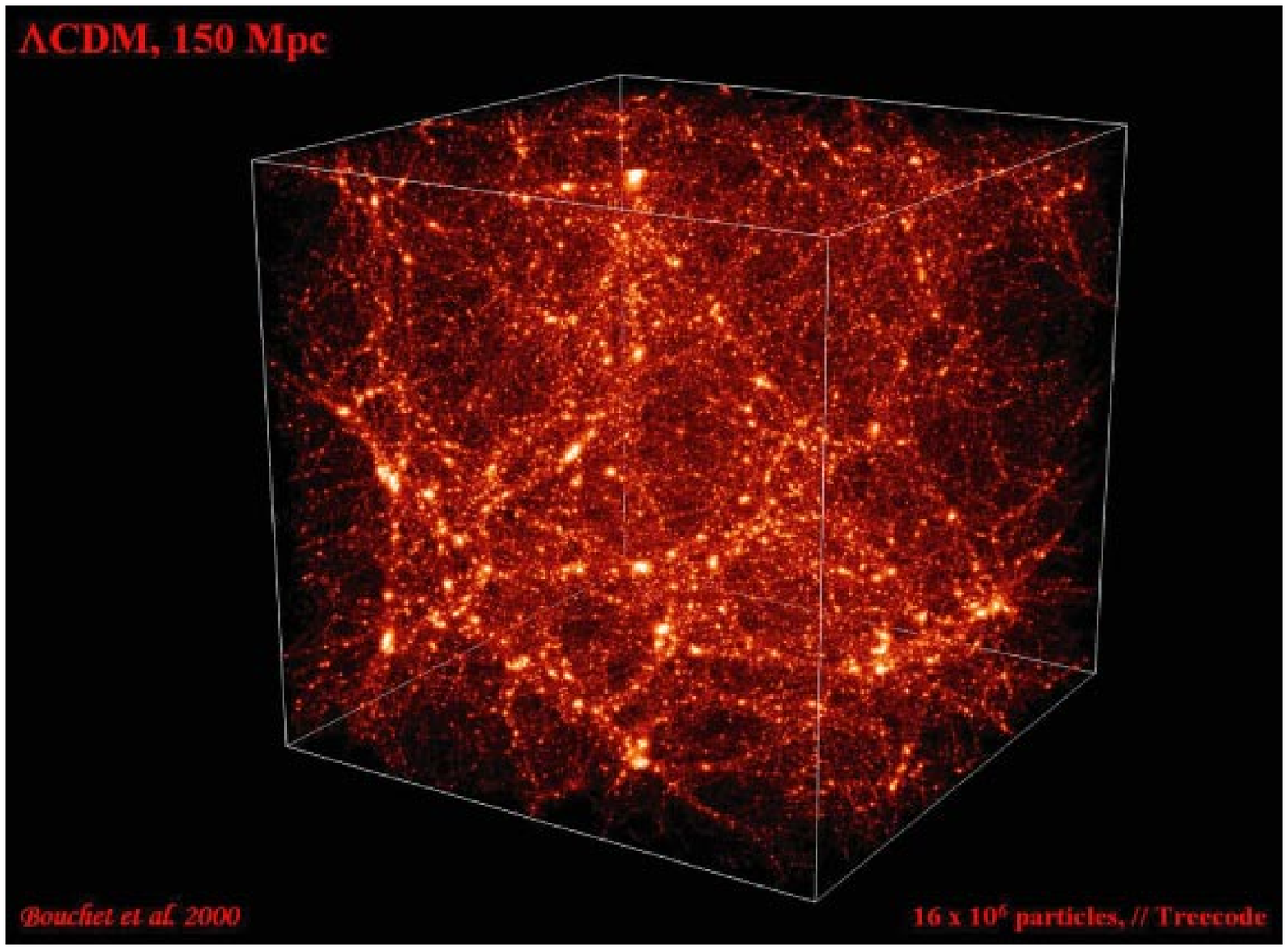, width=0.5\textwidth}
\psfig{file=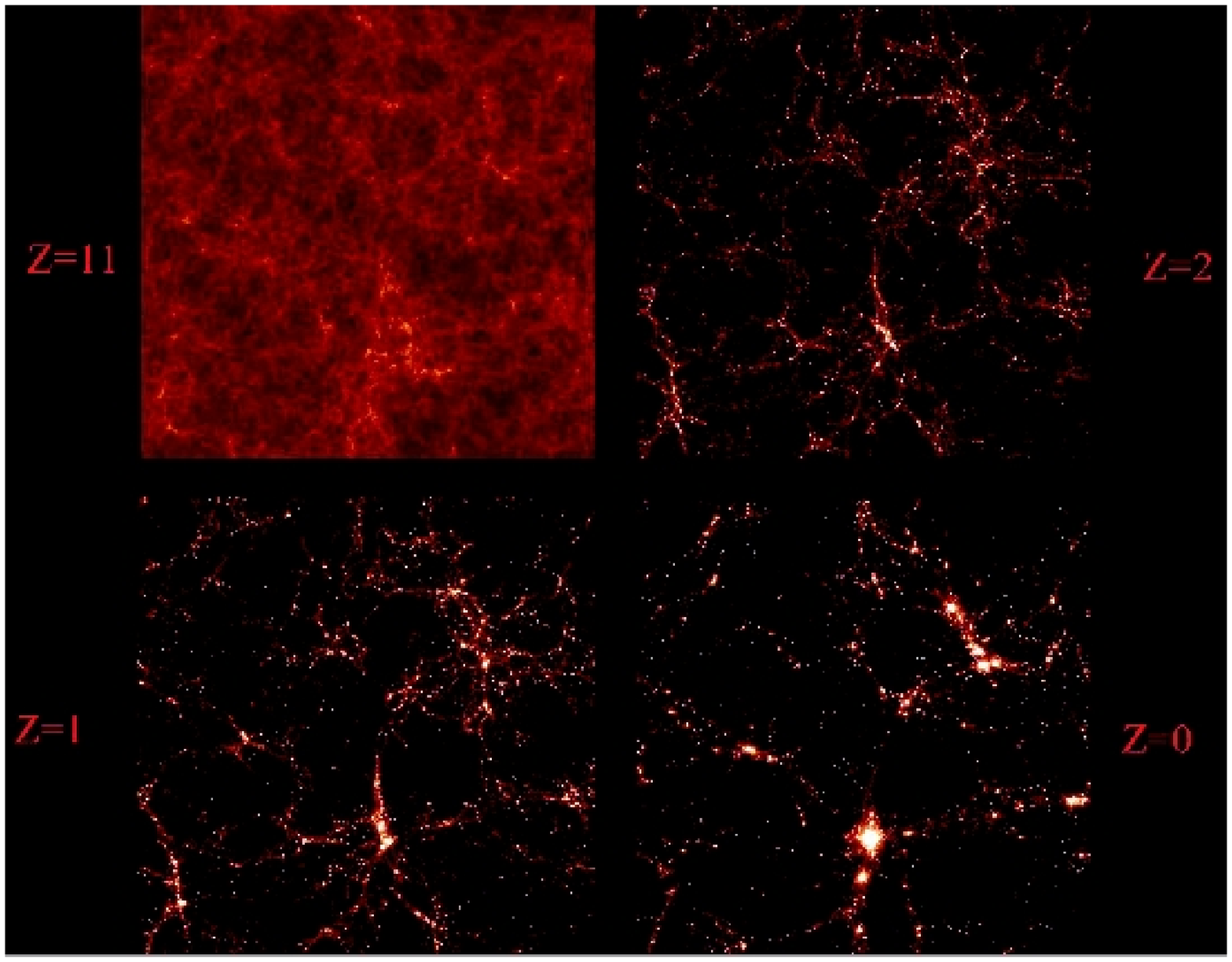, width=0.474\textwidth} }
\end{center} \vspace{-19pt}
\caption{A numerical simulation in a 150\,\Mpc\ box of a LCDM Universe
 ($\Omega = 1$, $\Omega_\Lambda =2/3$, $n_S =1$). a) (left) Resulting
 distribution of the CDM at present (luminosity proportional to the
 density). b) (right) Temporal evolution by gravitational instability of a
 thin (15\, Mpc) slice of the box showing the hierarchical development of
 structures within a global cosmic web of increasing contrast, but quite
 discernible early on.}
\label{fig:simus}
\end{figure}

Figure~\ref{fig:simus}.a shows the generated structures in the CDM components
in a numerical simulation box of 150\,\Mpc, while \ref{fig:simus}.b shows the
evolution with redshift of the density in a thin slice of that box. The
statistical properties of the derived distribution (with the cosmological
parameters given above) appear to provide a close match to those derived from
large galaxy surveys. Note that this simulation with $\Omega_\Lambda = 2/3,
\Omega_M=1/3$ was performed in 2000, well before the WMAP results. Indeed it
was already by then the favourite model.

When collapsed objects are formed, the baryonic gas initially follows the
infall. But shocks will heat that gas, which can later settle in a disk and
cool, and form stars and black holes which can then feed back through ionising
photons, winds, supernovae\ldots on the evolution of the remaining gas (after
first reionising the Universe at $z > 6$). In this picture, galaxies are
therefore (possibly biased) tracers of the underlying large scale structures
of the dark matter.

\section{Physics overview of CMB anisotropies\label{sec:basicphys}} 

In this standard cosmological model, processes in the very early universe
generate the seed fluctuations which ultimately give rise to all the
structures we see today. In the early universe, baryons and photons were
tightly coupled through Thomson scatterings of photons by free electrons (and
nuclei equilibrated collisionally with electrons). When the temperature in the
universe became smaller than about 3000 K (which is much lower than 13.6\, eV
due to the large number of photons per baryons $\sim 1.5\ 10^9$), the cosmic
plasma recombined and the ionisation rate $x_e $ fell from 1 at $z > 1100$
down to $x_e <10^{-3}$ at $z < 1100$: the photons mean free path $\propto
1/x_e$ rapidly became much larger than the Horizon $\sim c H^{-1}$. As a
result, the universe became transparent to background photons, over a narrow
redshift range of 200 or less. Photons then propagated freely as long as
galaxies and quasars did not reionise the universe (but by then the density
will have fallen enough that only a small fraction was rescattered). We
therefore observe a thin shell around us, the last scattering ``surface'' (LSS
in short) where the overwhelming majority of photons last interacted with
baryonic matter, at a redshift of 1100, when the Universe was less than 400
000 years old. The anisotropies of the CMB are therefore the imprint of the
fluctuations as they were at that time (but for a small correction due to the
photons propagation through the developing Large Scale Structures).

\begin{figure}[htb] \begin{center} \hbox{
\psfig{file=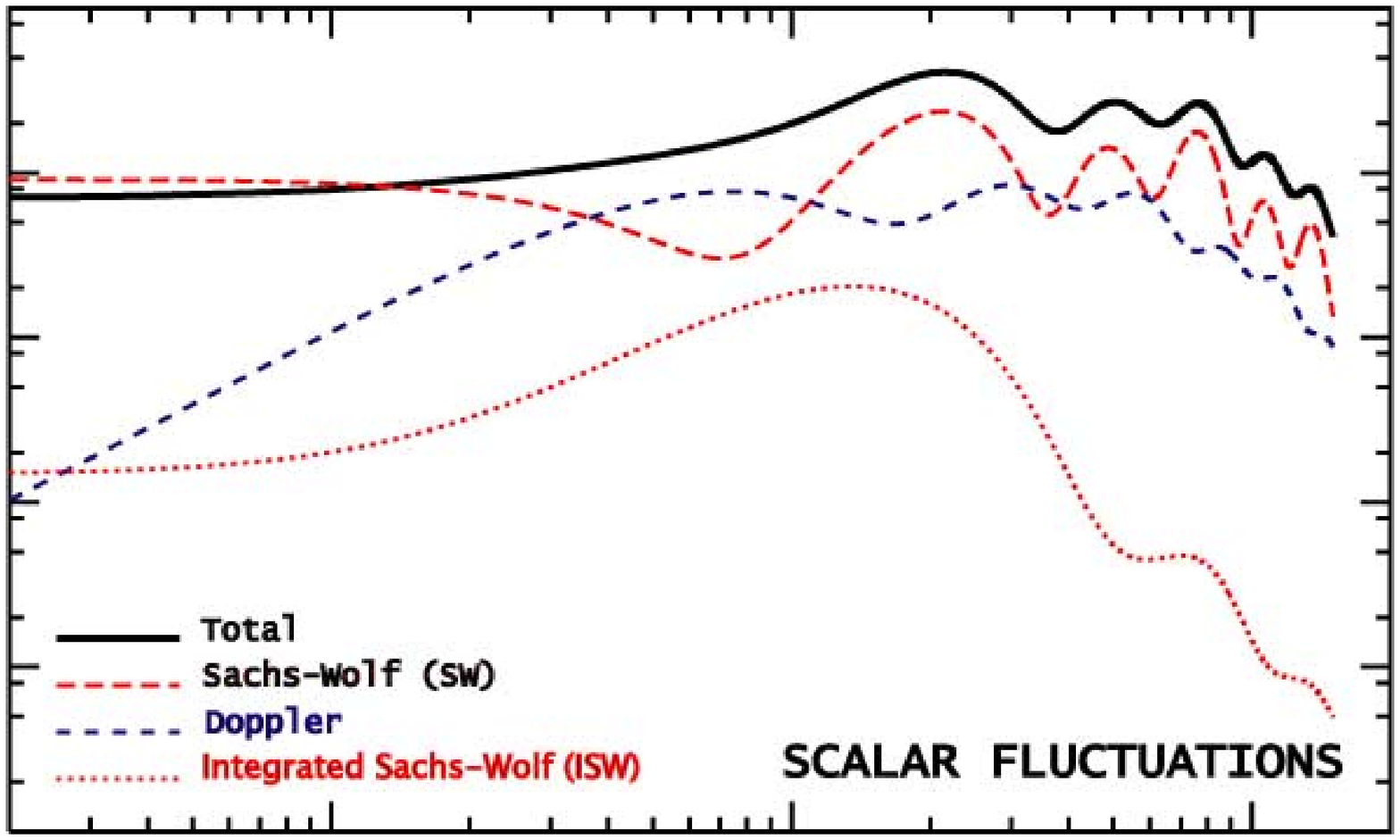, width=0.5\textwidth, height=60 truemm}
\psfig{file=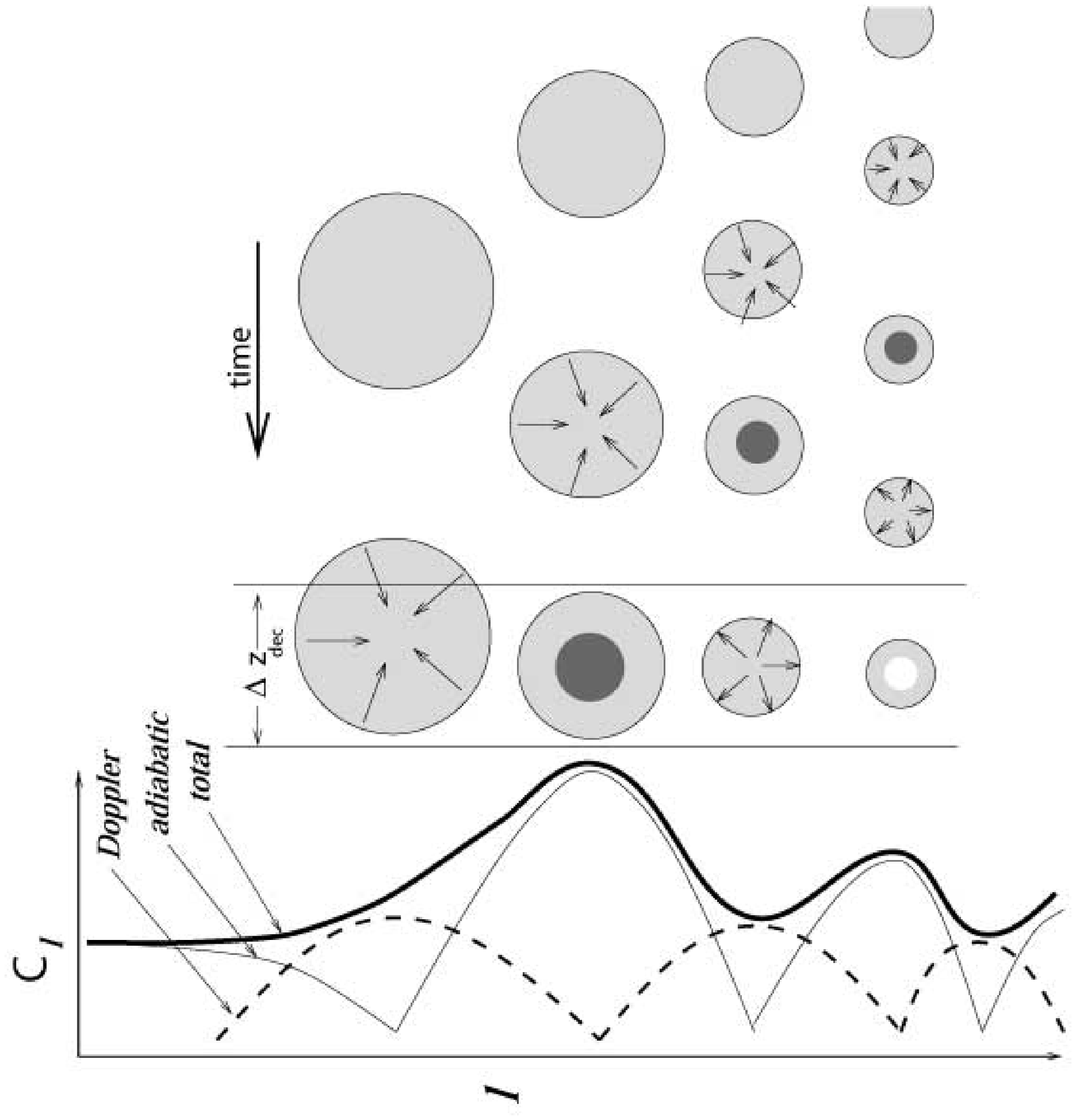, width=0.5\textwidth, height=60 truemm} }
\end{center} 
\vspace{-20pt}
\caption[]{The expected shape of the angular power spectrum of the temperature
  anisotropies, $C(\ell )$ (times $\ell^2$ to give the logarithmic
  contribution of each scale to the variance). a) Relative contributions; it
  has been assumed that only scalar fluctuations are present. The plot is in
  log-log coordinates. b) As time progresses, larger and larger fluctuations
  start oscillating and leave their characteristic imprint on the spectrum
  (reprinted from \cite{0305179}).}
\label{fig:cl_theo}
\end{figure}

To analyse the statistical properties of the temperature anisotropies, we can
either compute the angular correlation function of the temperature contrast
$\delta_T$, or the angular power spectrum $C(\ell)$ which is it's spherical
harmonics transform (in practice, one transforms the $\delta_T$ pattern in
$a(\ell, m)$ modes and sums over $m$ at each multipole since the pattern
should be isotropic, at least for the trivial topology, in the absence of
noise). A given multipole corresponds to an angular scale $\theta \sim
180^o/\ell$. These two-point statistics characterise completely a Gaussian
field. Figure~\ref{fig:cl_theo}.a shows the expected $C(\ell)$ shape in the
context we have described above.

This specific shape of the $C(\ell)$ arises from the interplay of several
phenomena. The most important is the so-called ``{\em Sachs-Wolf effect}''
\cite{SachsWolfe67} which is the energy loss of photons which must ``climb
out'' of gravitational potential wells at the LSS (to ultimately reach us to
be observed), an effect which superimposes to the intrinsic temperature
fluctuations (we therefore observe an effective temperature which sums this
effects, of opposite sign for adiabatic initial conditions where every
component [$\gamma$, baryons, CDM\ldots] is perturbed
simultaneously). Figure~\ref{fig:cl_theo}.b gives a pictorial view of the
temporal development of primordial fluctuations at different scales (top) and
how the state of fluctuations translates at recombination in the power
spectrum of CMB fluctuations (bottom).

\begin{figure}[htb] \begin{center}
\psfig{file=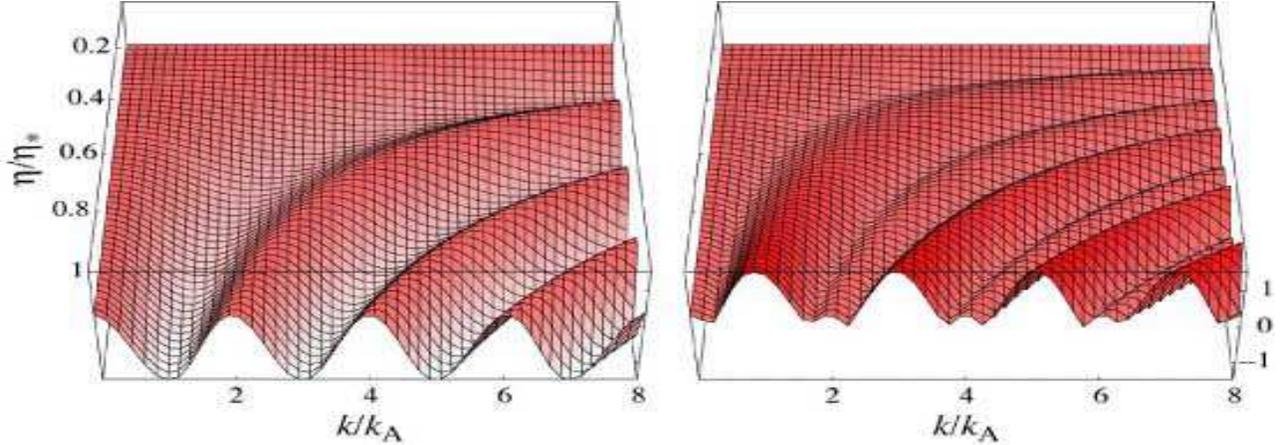, width=\textwidth, height= 60mm}
\end{center} \vspace{-20pt}
\caption[]{Temporal evolution of the effective temperature which sums the
  effects (of opposite sign) of the intrinsic temperature and of the Newtonian
  potential fluctuations (for $R=(p_B+\rho_B) / (p_\gamma + \rho_\gamma) =$
  cste). a) (left) Amplitude. Note the zero point displacement which leads to
  a relative enhancement of compressions. b) (right) rms showing the enhanced
  odd-numbered peaks. Reprinted from \cite{2003AnPhy.303..203H}.}
\label{fig:osc}
\end{figure}

Since the density contrasts of these (scalar) fluctuations is very weak, one
can perform a linear analysis and study each Fourier mode independently (the
effect of the primordial spectrum will thus simply be to weight the various
modes in the final $C(\ell)$). Figure~\ref{fig:osc} shows the (approximate)
temporal evolution of the amplitude of different Fourier modes. While gravity
tends to enhance the contrast, the (mostly photonic) pressure resists and at
some points stops the collapse which bounces back, and expands before
recollapsing again\ldots This leads to acoustic oscillations, on scales small
enough that the pressure can be effective, i.e. for $k > k_A$, where the
acoustic scale $k_A$ is set by the inverse of the distance travelled at the
sound speed at the time $\eta_\star$ considered. On scales larger than the
sound horizon ($k < k_A \propto 1/(c_S\, \eta_\star)$), the initial contrast
is simply amplified. At $k =k_A$ the amplification is maximal, while at $k=2
k_A$ it had time to fully bounce back. More generally, the odd-multiples of
$k_A$ are at maximal compression, while it is the opposite for the even
multiples of $k_A$.

One should note the displacement of the zero point of the oscillations which
results from the inertia that baryons bring to the fluid. The rms of the modes
amplitude (right plot) therefore show a relative enhancement of the odd
(compression) peaks versus the even (rarefaction) ones, this enhancement being
directly proportional to the quantity of baryons, i.e. $\Omega_B h^2$, if $h$
stands for the Hubble ``constant'', $H = \dot a / a$, in units of 100 km/s/Mpc
(today $h_0 = 0.72 \pm 0.05$). Note that since $\Omega_X$ stands for the ratio of
the density of $X$ to the critical density (such that the Universe is
spatially flat), and since that critical density decrease with time as
$h^{-2}$, $\Omega_X h^2$ is indeed proportional to the physical density of
$X$.

Let us assume that the LSS transition from opaque to transparent is
instantaneous, at $\eta = \eta_\star$. What we would see then should just be
the direct image of these standing waves on the LSS; one therefore expects a
series of peak at multipoles $\ell_A = k_A \times  D_\star$, where $D_\star$ is the
angular distance to the LSS, which depends on the geometry of the
spacetime. Of course a given $k$ contributes to some range in $\ell$ (when $k$
is perpendicular to the line of sight, it contributes to lower $\ell$ than
when it is not), but this smearing is rather limited.  The dependence of the
acoustic angular scale $\ell_A$ on geometry and sound speed leads to its
dependence on the values of three cosmological parameters. One finds for
instance
\begin{equation} 
\frac{\Delta\ell_A}{\ell_A} \simeq -1.1 \frac{\Delta\Omega}{\Omega} -0.24
\frac{\Delta\Omega_M h^2}{\Omega_M h^2} + 0.07 \frac{\Delta\Omega_B
h^2}{\Omega_B h^2}
\label{eq:la}\end{equation} 
around a flat model $\Omega=1$ with 15\% of matter ($\Omega_M h^2 =0.15$) and
2\% of baryons \cite{2003AnPhy.303..203H}. Note that this information on the
peaks positions (and in particular that of the first one) is mostly dependent
on the total value of $\Omega$ (geometry), with some weaker dependence on the
matter content $\Omega_M h^2$, and an even weaker one on the baryonic density.

Concerning the latter, as already mentioned, baryons increase the inertia of
the baryon-photon fluid and shifts the zero point of the oscillations. A
larger baryonic density tends to increase the contrast between odd and even
peaks; one can therefore use this contrast-in-height information as a
baryometer.  The influence of dark matter is more indirect. By increasing its
quantity, one increases the total matter density and advances the time when
matter comes to dominate the energy density of the universe. This changes the
duration of time spent in the radiation and matter dominated phase, who may
have different growth rates.
The net effect is to globally { decrease} the first peaks amplitude when the
matter content increases (in addition to the small shift in scale due to the
variation of $\ell_A$ already noted above). The effect on the power spectrum
peaks of all the matter is thus rather different from that of the baryonic
component alone. Therefore the shape of the spectrum is sensitive to both
separately. This suggests that degeneracies in the effect of these three
parameters ($\Omega, \Omega_M h^2, \Omega_B h^2$) can be lifted with
sufficiently accurate CMB measurements, a statement which more detailed
analyses confirm.

While the reader in a hurry can skip to the conclusion of this section, I now
turn to describing other effects which must be taken into account to fully
understand the shape of the power spectrum.  Since the fluid is oscillating,
there is also a {\em Doppler effect} in the $k$ direction (blue dotted line in
figure~\ref{fig:cl_theo}.a, which is zero at the acoustic peaks and maximal in
between. This effect adds in quadrature to the Sachs-Wolf effect considered so
far. Indeed, imagine an acoustic wave with $k$ perpendicular to the line of
sight, we see no Doppler effect, while for a $k$ parallel to the line of
sight, the Doppler effect is maximal and the Sachs-Wolf effect is null. This
smoothes out the peak and trough structure, although not completely since the
Doppler effect is somewhat weaker than the SW effect (by an amount $\propto
\Omega_B h^2$). Therefore an increase in the baryonic abundance also increases
the peak-trough contrast (in addition to the odd-even peak contrast).

So far we considered the fluid as perfect and the transition to transparence
as instantaneous, none of which is exactly true. Photons scattered by
electrons through Thomson scattering in the baryons-photons fluid perform a
random walk and diffuse away proportionally to the square root of time (in
comoving coordinates which remove the effect of expansion). Being much more
numerous than the electrons by a factor of a few billion, they drag the
electrons with them (which by collisions drag in turn the protons). Therefore
all fluctuations smaller than the diffusion scale are severely damped. This
so-called Silk damping is enhanced by the rapid increase of this diffusion
scale during the rapid but not instantaneous combination of electrons and
protons which leads to the transparence. As a result of the finite thickness
of the LSS and the imperfection of the fluid, there is {\em an exponential
cut-off of the large-$\ell$ part} of the angular power spectrum. As a result,
there is not much primordial pattern to observe at scales smaller than $\sim
5^\prime$.

After recombination, photons must travel through the developing large scale
structures to reach the observer. They can lose energy by having to climb out
of potential wells which are deeper than when they fell in (depending on the
rate of growth of structures, which depends in turn on the cosmological
census). Of course the reciprocal is also true, i.e. they can gain energy from
forming voids. These tend to cancel at small scale since the observer only
sees the integrated effect along the line of sight. The red dotted line of
figure~\ref{fig:cl_theo} shows the typical shape of that {\em Integrated
Sachs-Wolf} (ISW) contribution. The ISW is anti-correlated with the Sachs-Wolf
effect, so that the total power spectrum $C(\ell)$ is in fact a bit smaller
than the sum of each spectrum taken separately. Finally, other small secondary
fluctuations might also leave their imprint, like the lensing of the LSS
pattern by the intervening structures, which smoothes slightly the spectrum.
But that can be predicted accurately too. And in fact the smoothing kernel
dependence on cosmological parameters introduces small effects that may help
reducing some residual degeneracies between the effect of parameters on the
power spectrum shape.

Other secondary effects, imprinted after recombination, are generally much
weaker (at scales $> 5\prime$). For instance, the {\em Rees-Sciama} effect
\cite{ReesSciama68} (a non-linear version of the ISW) generates temperature
fluctuations, with amplitudes of about a few $10^{-7}$ to $10^{-6}$; its
amplitude is maximum for scales between 10 and 40 arc minutes
\cite{1996ApJ...460..549S}). At the degree scale, this contribution is only of
the order of 0.01 to 0.1\% of the primary CMB power. The inverse Compton
scattering of the CMB photons on the free electrons of hot intra-cluster gas
produces the {\em Sunyaev \& Zeldovich} (SZ) effect \cite{ZeldovichSunyaev69,
1972A&A....20..189S, 1980MNRAS.190..413S}). This effect has a specific
spectral signature which should allow separating it, at least in sufficiently
sensitive multi-frequency experiment. But the motion along the line of sight
of clusters induces a first order Doppler effect, usually called the {\em
Kinetic Sunyaev-Zeldovich} effect, which is a true source of temperature
fluctuation, albeit rather weak (the rms cluster velocity is $\sim 10^{-3} c$)
and in the specific direction of clusters. A similar effect, the {\it
Ostriker-Vishniac effect} \cite{1986ApJ...306L..51O, 1987ApJ...322..597V}
arises from the correlations of the density and velocity perturbations along
the line of sight, when the universe is totally ionised. The corresponding
anisotropies are at the few arc-minute scale and their amplitudes depend much
on the ionisation history of the universe \cite{1995ApJ...439..503D,
1996A&A...315...33H}). However, they remain smaller than the primary
anisotropies for $\ell < 2000$. This type of Doppler effect can in fact happen
in all sorts of objects containing ionised gas, like expanding shells around
the sources that reionised the Universe \cite{1996A&A...311....1A,
1998ApJ...508..435G, KnoxScoDod.prl, 1999A&A...341..640A}), or primordial
galaxies hosting super-massive black holes \cite{2000A&A...357....1A}, but the
relevant angular scales are rather in the arc second range or smaller. This
does not hold true however for various foreground emission, like those of our
own Galaxy. But like for the Sunyaev-Zeldovich effect, one can use
multi-frequency observations to separate them out rather well.

In summary, {\em the seeds of large scale structures must have left an imprint
on the CMB, and the statistical characteristics of that imprint can be
precisely predicted as a function of the properties of the primordial
fluctuations and of the homogeneous Universe}. Reciprocally, we can use
measurements of the anisotropies to constrain those properties.

\section{Observations of CMB anisotropies\label{sec:observ}}

Many location have been considered for CMB observations: ground based
telescopes at many places in the world, included at South Pole, air-planes,
stratospheric balloons, satellites around the Earth, and satellites far from
the Earth. The main parameters related to the location are the atmospheric
absorption and emission and the proximity of sources of straylight.  The Earth
atmosphere is a very strong source at millimetre and submillimetre
wavelengths~\cite{PajotThese}. It is therefore a source of photon noise. In
addition, it is not perfectly uniform and the slowly changing structure of its
emission adds noise that can be confused with CMB anisotropies. Even at
stratospheric balloon altitudes (38 to 40 km), one may encounter structures at
low angular frequencies that limit measurement stability. Only satellites are
free of this first ``layer'' of spurious emission.

Straylight coming from the Earth is a common problem for all locations except
for space probes distant from the Earth as seen from the instrument. The
brightness temperature of about 250K of the Earth times its solid angle has to
be compared with the instrument sensitivity in brightness times the beam solid
angle. For a mission in low earth orbit the ratio is $6 \times 10^{13}$ and at the
L2 Lagrange point the ratio is $5 \times 10^9$. At the wavelengths of interest,
diffraction is a major source of straylight, and modelling and experimentation
can hardly estimate reliably or measure the very low side lobes needed to meet
this extremely high rejection ratio. The problem is severe enough to have
motivated the choice of the Lagrangian point L2 of the Sun-Earth system, at
about 1.5 million kilometres from Earth, to locate the new generation of
satellites W\maps and \planck.

\begin{figure}[htb] 
\begin{center} \vbox{ \hbox{ 
\psfig{file=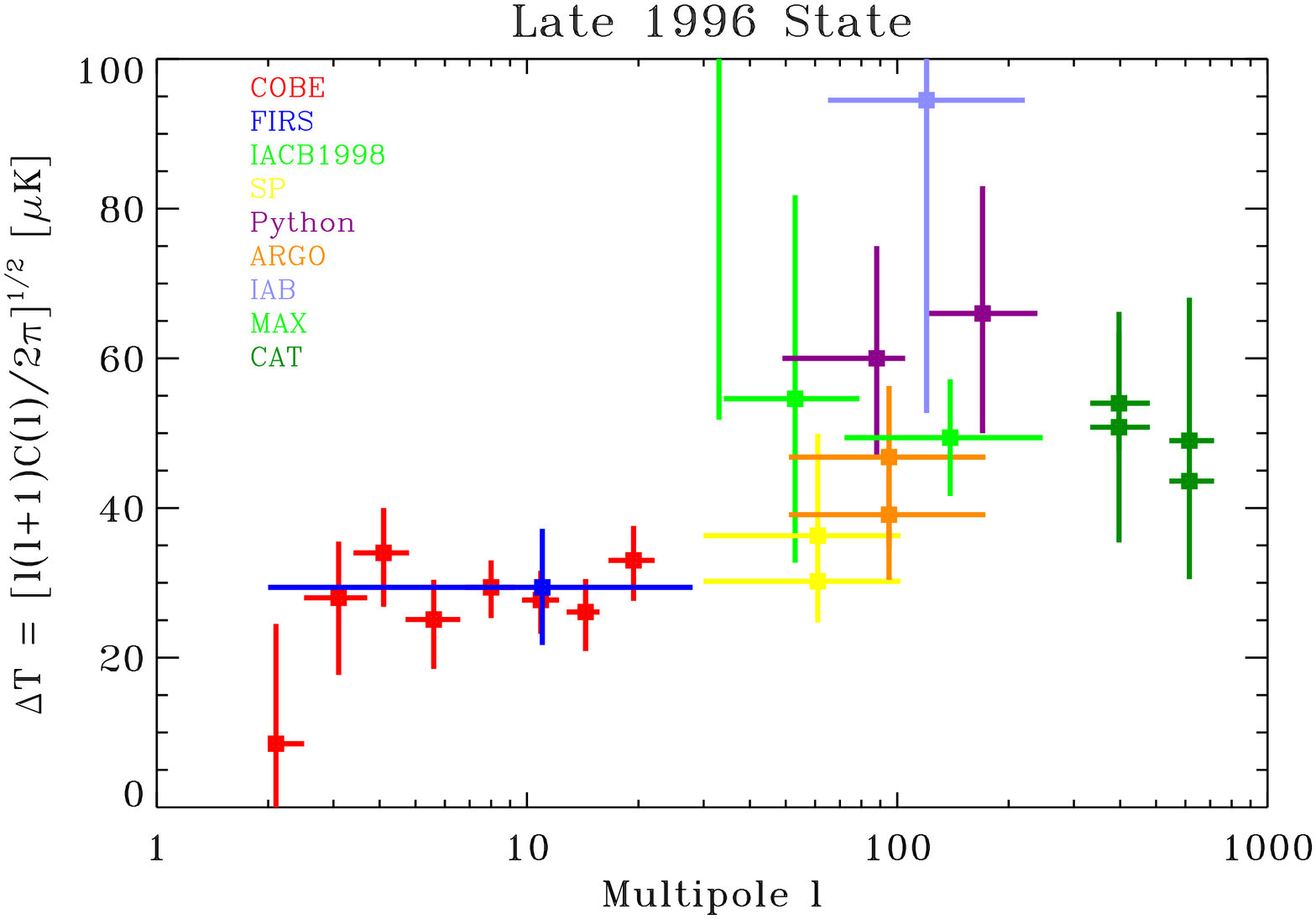, angle=90, width=0.5\textwidth} \psfig{file=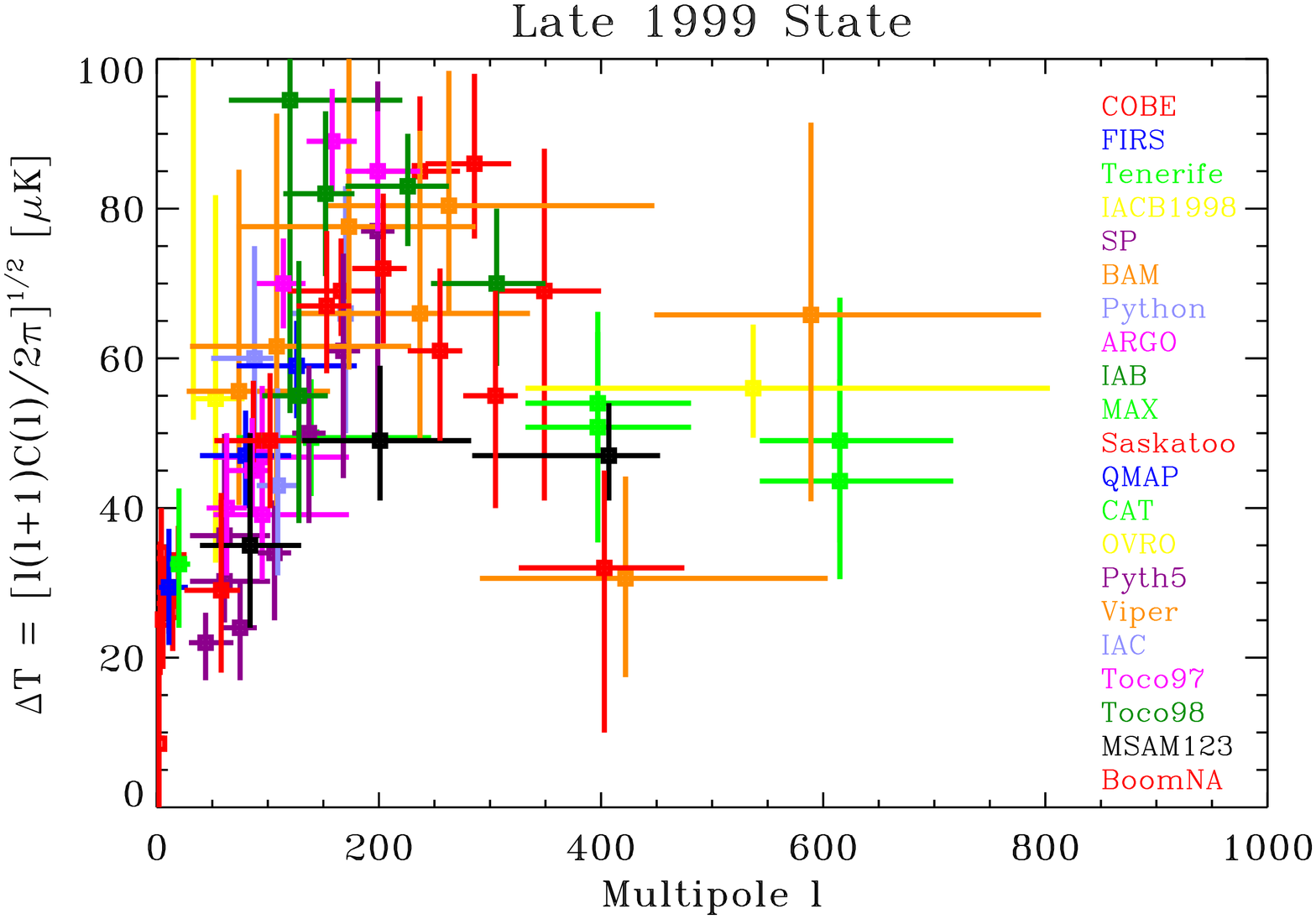,
angle=90, width=0.5\textwidth} } \hbox{ \psfig{file=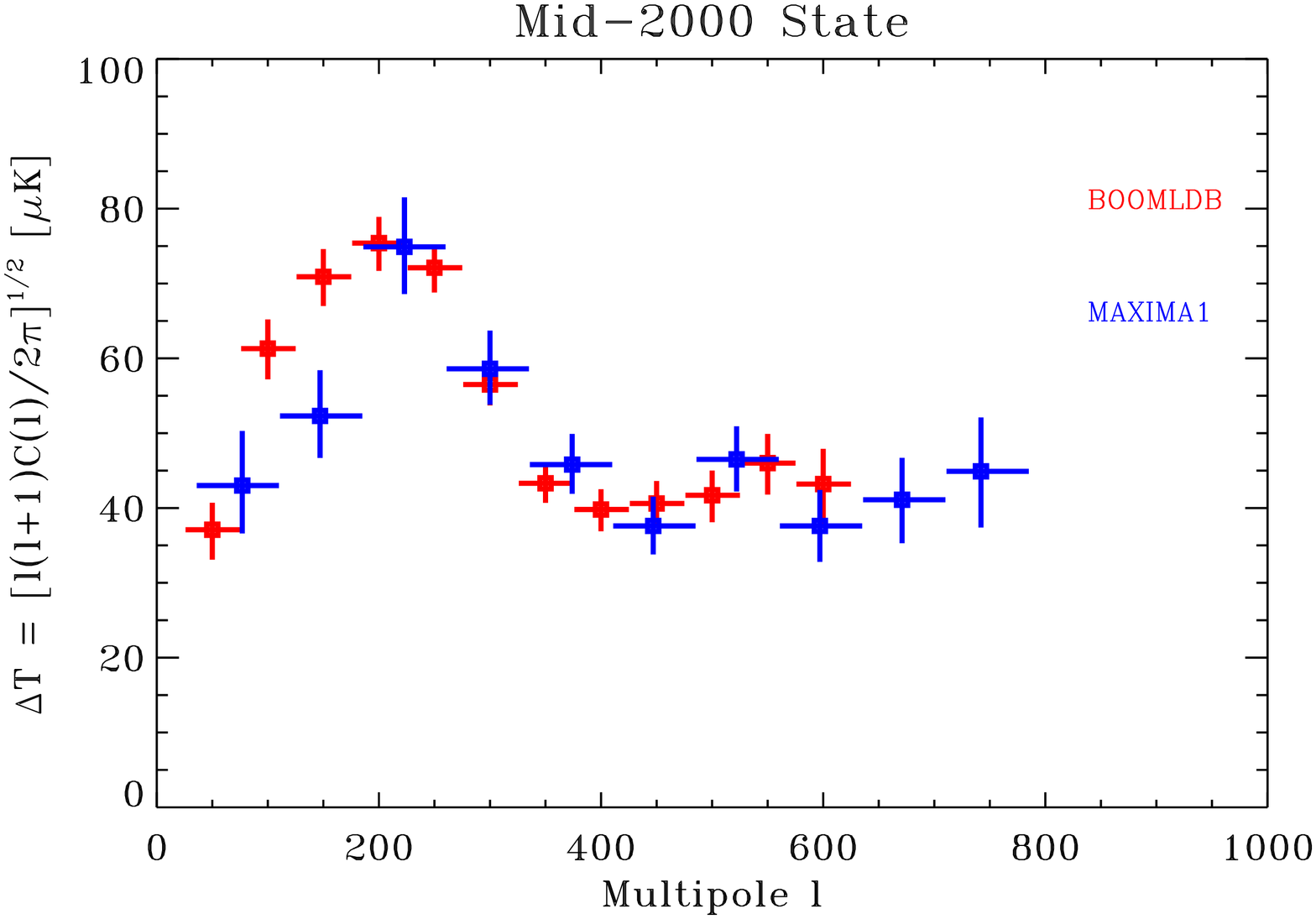,
angle=90,width=0.5\textwidth}
\psfig{file=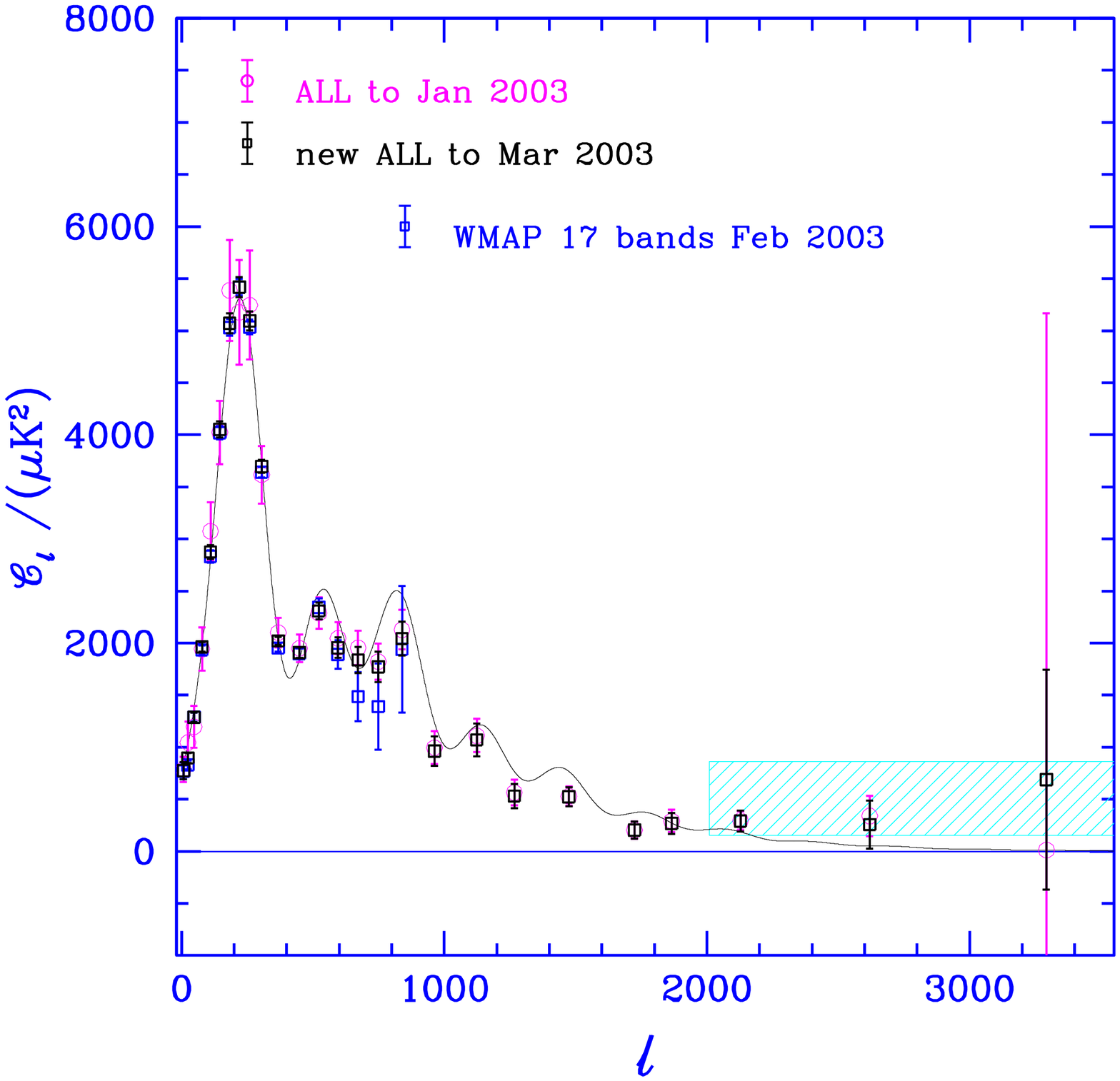,
width=0.5\textwidth, height= 55 truemm} } }
\end{center} \vspace{-10pt}
\caption[Measurements of the power spectrum]{Successive Measurements of the
  power spectrum. The first panel at top left shows all published detection at
  the end of 1996, while the second plot at right is an update at the end of
  1999. The left bottom plot shows the two results published in may 2000 by
  the BOOMERanG and MAXIMA teams (with each curve moved by + or - 1 $\sigma$
  of their respective calibration). The final panel at bottom left (courtesy
  D. Bond) shows in purple the optimal spectra in 26 bands (allowing for
  calibration errors in each experiment) from the co-analysis of all data
  (including Archeops, the extended VSA, ACBAR and a preliminary version of
  the 2 year CBI data) till January 2003, as well as the WMAP spectrum (blue
  points), and the spectrum as we know it today (black points) when all
  experimental evidence are used.}
\label{fig:history_cl}
\end{figure}

The first clear detection of the CMB anisotropies was made in 1992 by the DMR
experiment aboard the COBE satellite orbiting the earth with the DMR
instrument (and soon afterwards by FIRS), with a ten degree (effective) beam
and a signal to noise per pixel around 1.  This lead to a clear detection of
the large scale, low-$\ell$, Sachs-Wolf effect, the flatness of the curve (see
fig.~\ref{fig:history_cl}.a) indicating that the logarithmic slope of the
primordial power spectrum, $n_S$, could not be far from one. The $\sim 30\muK$
height of the plateau gave a direct estimate of the normalisation of the
spectrum, $A_S$ (assuming the simplest theoretical framework, without much
possible direct checks of the other predictions given the data)

In the next four years (fig~\ref{fig:history_cl}.a), a number of experiments
started to suggest an increase of power around the degree scale, i.e. at $\ell
\sim 200$. As shown by fig~\ref{fig:history_cl}.b, by 1999 there was clear
indication by many experiments taken together that a first peak had been
detected. But neither the height nor the location of that peak could be
determined precisely, in particular in view of the relative calibration
uncertainties (and possible residual systematics errors).

That situation changed in may 2000 when the BOOMERanG and Maxima
collaborations both announced a rather precise detection of the power spectrum
from $\ell \sim 50$ to $\ell \simgt 600$. That brought a clear determination
of the first peak around an $\ell$ of 220 (see panel c), with the immediate
implications that $\Omega$ had to be close to one. This result had
considerable resonance since it clearly indicated, after decades of intensive
work, that the spatial geometry of the Universe is close to flat, with of
course the imprecision due to the poor determination of the other parameters
which also have an influence, albeit weaker, on the position of that peak (see
eq.~\ref{eq:la}).

\begin{figure}[htb] \begin{center}
\psfig{file=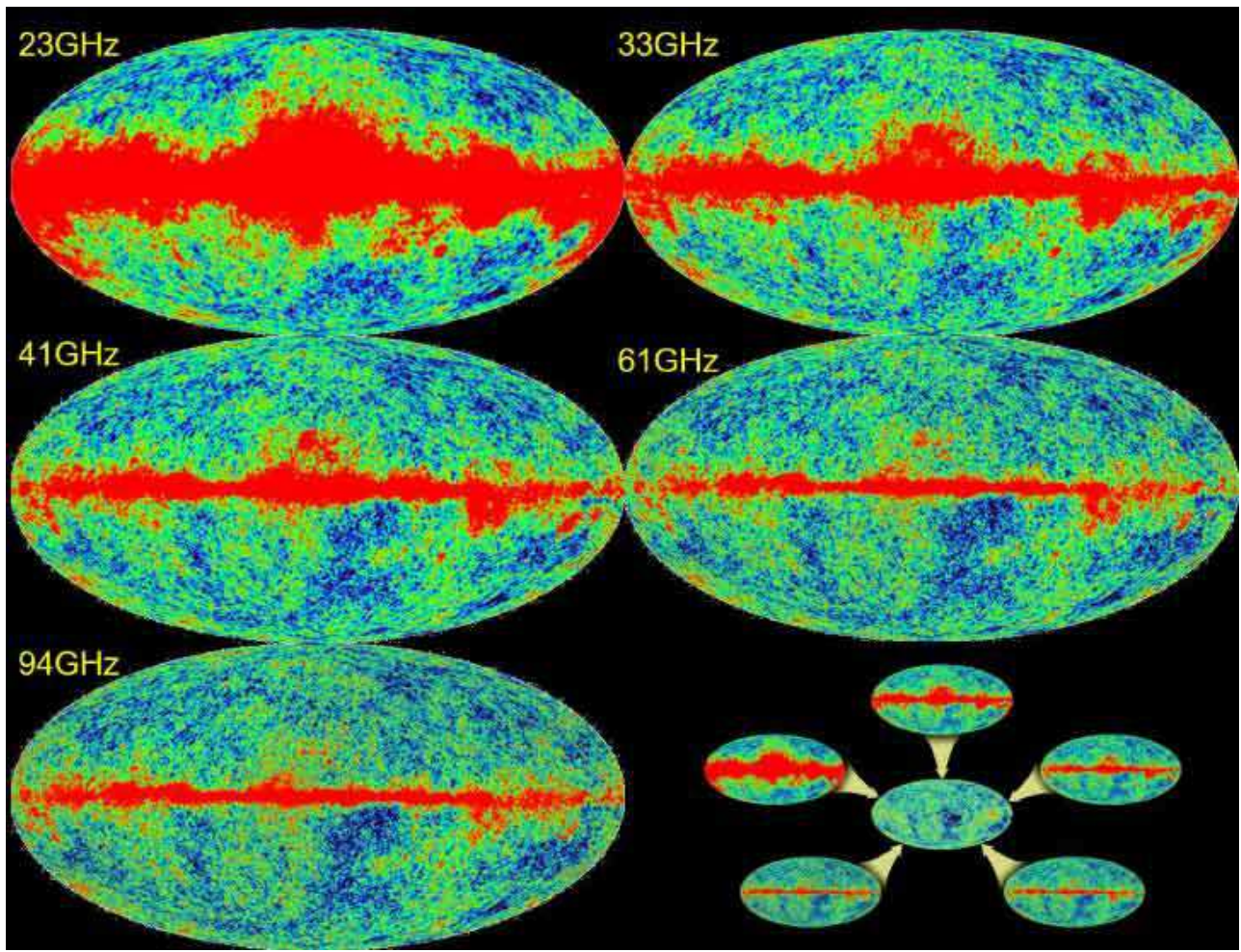, width=\textwidth} \end{center}
\vspace{-10pt}
\caption[]{The WMAP maps at each frequency, which shows in particular the
  varying strength of the Galactic emissions according to frequency, and in
  the bottom right an icon illustrating that these informations are merged to
  extract an estimate of the CMB part.}
\label{fig:wmap_maps}
\end{figure}

As recalled earlier, a crucial prediction of the simplest adiabatic scenario
is the existence of a series of acoustic peaks whose relative contrast between
the odd and even ones gives a rather direct handle on the baryonic
abundance. In addition, one expects to see at larger $\ell$ the damping
tail. All of these have now been established by the DASI 2001 experiment, an
improved analysis of BOOMERanG, and foremost by the release in may of 2002 of
the VSA and CBI results. In addition the Archeops experiment gave at the end
of 2002 a quite precise determination of the low-$\ell$ part of the
spectrum. Panel d of fig~\ref{fig:history_cl} shows a co-analysis performed by
D. Bond of all results obtained till the end of 2002, as well as the recent
determination by the WMAP satellite. Clearly all the pre-WMAP ground
experiments had done quite a wonderful job at pinning down the shape of the
temperature spectra. This panel also shows the spectrum as we know it today,
when all experimental results are analysed together.

\begin{figure}[htb] 
\begin{center}
\psfig{file=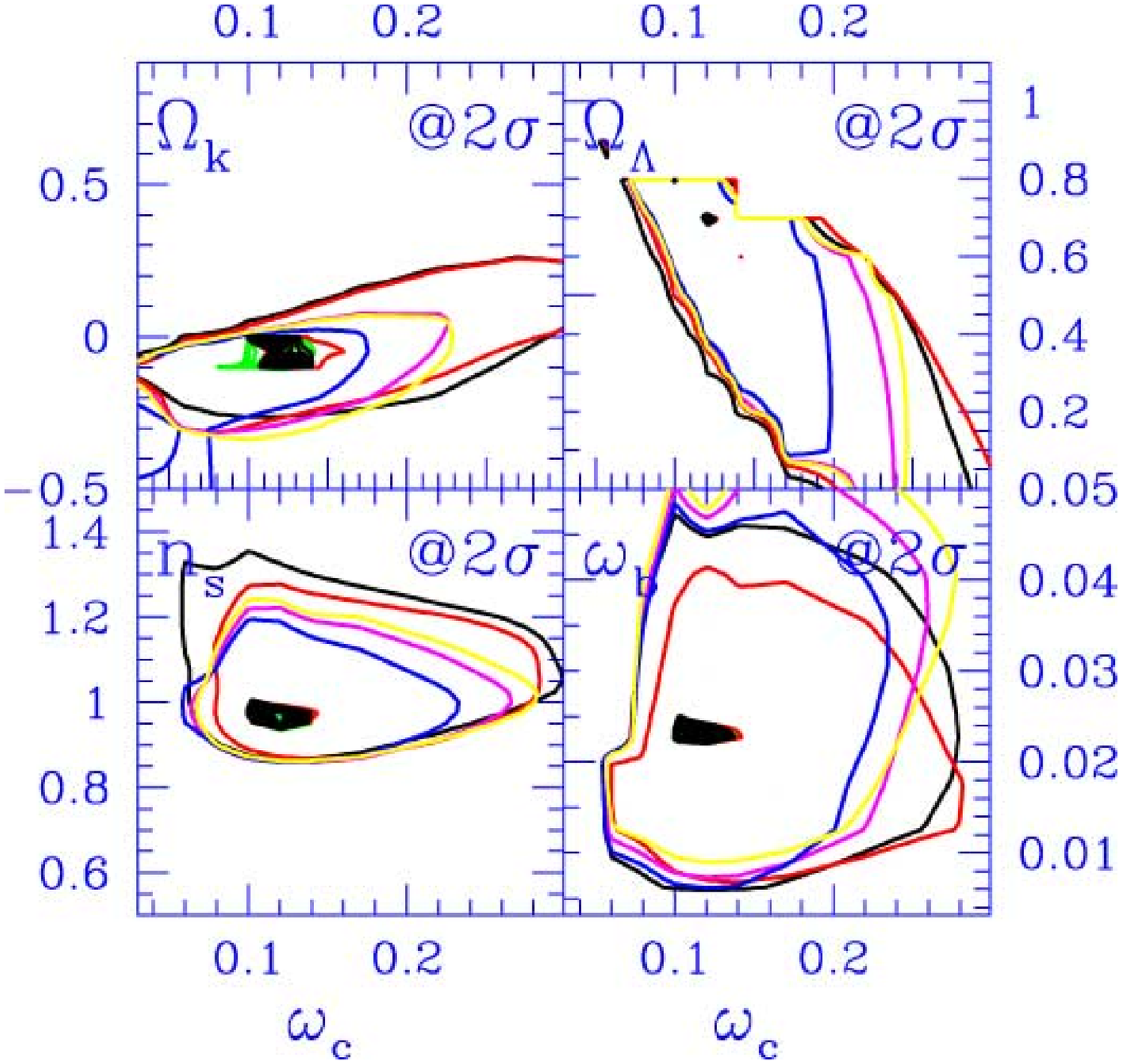,
width=\textwidth, height=0.5\textwidth}
\end{center} \vspace{-10pt}
\caption[]{Successive constraints in the $\Omega_k =1-\Omega, \Omega_\Lambda,
 n_S, \omega_B = \Omega_B h^2$ versus $\omega_c=\Omega_{CDM} h^2$ cuts in the
 global parameter space fitted to the $C(\ell)$ successive data, using
 COBE/DMR in all cases. The colour coding is the following: CBI = black, VSA =
 (outer) red, Archeops = yellow, ACBAR = blue, Ruhl cut for Boomerang =
 magenta. Green corresponds to Acbar + Archeops + Ruhl + DASI + Maxima + VSA +
 CBI. The interior red is all of the above + WMAP with no prior on $\tau$ ,
 while black is with a $tau$ prior motivated by their ``model independent''
 result from the TE analysis (it is broader than a $0.16 \pm 0.04$
 Gaussian). (Courtesy D. Bond) }
\label{fig:params}
\end{figure}

The figure~\ref{fig:params} shows the constraints successively posed by these
CMB experiments on some of the parameters of the model, using only weak priors
arising from other cosmological studies. This priors state that the current
Hubble ``constant'', $H_0 = 100 h^{-1}$km/s/Mpc, has to have a value between
45 and 90 km/s/Mpc, that the age of the Universe has to be greater than 10
Billion years and that the matter density is larger than 1/10 of the critical
density, all of which can be considered as very well established (if for
instance the Universe has to be older than it's oldest stars!).

The top left panel shows that indeed that the curvature term $\Omega_k =1 -
\Omega$ has to be close to zero. The panel on the right shows that
$\Omega_\Lambda$ and $\omega_c = \Omega_{CDM} h^2$ are not well determined
independently of each other by single experiments. This simply reflects the
fact that the $C(\ell)$ global pattern scales by the angular distance (recall
$\ell_A = k_A D_\star$) which is determined by the geometry (i.e. $\Omega =
\Omega_\Lambda + \Omega_{CDM}\ (+ \Omega_B)$), while the data is not precise
enough to uncover the subtler effects (sound speed, lensing\ldots) which break
that degeneracy. But this degeneracy was lifted by the co-analysis, even
before WMAP, and independently of the supernovae result\ldots

The bottom left panel lends support to the $n_s =1$ hypothesis. Many
inflationary models suggest value of $n_s$ slightly lower than one (and even
departures from a pure power law), but the data is not yet good enough to
address these questions convincingly. Completing the census, the bottom right
shows the contours in the $\omega_b$ - $\omega_c$ plane. The CMB determination
turns out to be in excellent agreement with the constraints from primordial
nucleosynthesis calculations which yield $\omega_b = \Omega_B h^2 = 0.019
\pm 0.002$.

In summary, this shows that many of the theoretical predictions corresponding
to the simplest scenario for the generation of initial conditions (Gaussian
statistics, adiabatic modes, no tensorial contribution, a scale invariant
power spectrum) in a flat Universe dominated by dark energy and cold dark
matter have now been detected, from the Sachs-Wolf plateau, to the series of
peaks starting at $\ell \simeq 220$, to the damping tail, together with a
first detection of the CMB polarisation at the expected level. The derived
parameters are consistent with the various constraints from other cosmological
probes and there are no glaring signs of inconsistencies. This was in fact in
place before the WMAP results, but it is remarkable that adding WMAP
essentially zooms-in onto the expected values, bringing now further support to
the model and its parameters, as recalled in {\S}~\ref{sec:struct}.


\section{Perspectives\label{sec:beyond}}

Cosmology therefore has a great {\em concordance} model, based on a minimal 7
parameters set ($\Omega, \Omega_{CDM},$ $\Omega_B, H, \tau$ to describe the
global evolution, and $n_S, A_S$ to describe the initial conditions) which
fits quite well all (multiple and partially independent) observational
evidence. But it might be too early to consider we have definitively
established the {\em standard} cosmological model, corroborated by many
independent probes.

In order to check the strength of the edifice, let us for instance consider
what it takes to stick to a simpler Einstein-de Sitter model (EdS, with
$\Omega_{Matter}=1, \Omega_\Lambda =0$) with no dark energy. As a matter of
fact, Blanchard et al. find in \cite{0304237} a great fit to the currently
measured $C(\ell)$ and a quite reasonable one to the $P(k)$, provided they
assume that 1) initial conditions are not scale invariant (the authors
consider an initial spectrum with 2 slopes) 2) there exists a non-clustering
component $\Omega_x=0.12$, both of which are quite possible. The Supernovae
data would then be the only source of independent evidence for a non-zero
$\Omega_\Lambda$, and many would question accepting the existence of such a
theoretically-unsettling component on that basis.  But this particular
Einstein de Sitter model does require also that $H_0 \sim 46$ km/s/Mpc. This
would require in turn that the HST measurement of $H_0 = 72 \pm  8$ km/s/Mpc be
completely dominated by an as-yet unknown systematics effect, which most
cosmologists are not ready to accept easily. Accepting this model would thus
require discarding two independent line of evidence.

In addition we may now have some rather direct evidence of the presence of
dark energy (meaning here $\Omega_\Lambda \ne 0$). Indeed, Bough \& Crittenden
\cite{0305001} and Fosalba \& Gaztanaga \cite{0305468} found a significant
cross-correlation between the CMB anisotropies measured by WMAP and various
tracers of Large Scale Structures. This was anticipated since the evolution of
potential wells associated to developing large scale structures when CMB
photons travel through them generally leaves an imprint, which we already
described, the Integrated Sachs-Wold effect (ISW). This imprint must of course
be correlated with tracers of LSS. It is interesting to note that, in the EdS
model, gravitationnal potential wells are linearly conserved in the matter
era, in which case no correlation should be found. Instead Bough \& Crittenden
\cite{0305001} found a non-zero value at the 2.4 to 2.8$\sigma$ level at zero
lag of the cross-correlation function of the WMAP map data with the hard X-Ray
background measured by the HEAO-1 satellite. They also found a somewhat less
significant correlation (at the 1.8 to 2.3$\sigma$ level) with an independent
tracer, the radio counts from the NVSS catalogue. Fosalba \& Gaztanaga
\cite{0305468} used instead the APM galaxy catalogue to build the (projected)
density field by smoothing at 5.0 \& 0.7 degree resolution and they also found
a substantial cross-correlation\ldots While these detections may not yet be at
a satisfactory level of significance for such an important implication, it
does bring a third line of evidence against the model proposed in
\cite{0304237}.

Let us therefore assume that this concordance model offers at least a good
first order description of the Universe. Still, deviations from this minimal
description remain quite possible and interesting.  One possibility much
debated recently is that what appears to be the manifestation of a
cosmological constant be rather a dynamical entity, for instance a
quintessence field with an equation of state where the pressure to density
ration is equal to $w(z)$ (the cosmological constant corresponding to the case
$w=-1$). Another possibility concerns a small contribution from massive
neutrinos. In both cases better CMB data might help determine these
effects. But the domain where most progress is expected and eagerly awaited
concerns the characterisation of initial conditions and its implications for
physics of the Early Universe. I now turn to a brief overview of various type
of deviations that future CMB observations might unearth or constrain much
more strongly.

Till now, we only considered {\em scalar} (adiabatic) fluctuations. Vectorial
perturbations tend to decay with expansion and are not predicted to leave any
imprint on the spectrum. But it is anticipated that that the very same process
that generated the primordial scalar fluctuations also created a stochastic
background of gravity waves. This weak tensorial contribution only appears in
the low-$\ell$ part of the temperature anisotropies spectrum, before the first
peak (since these waves decay as soon as there wavelength becomes smaller than
the horizon, which roughly corresponds in $\ell$ to the first peak). At the
current level of experimental precision it is most often ignored. But this
contribution is (relatively) much more significant for the polarised part of
the emission to which we now turn.

Thomson scatterings can create polarisation provided the incident radiation is
not isotropic, which can be induced by velocity gradients in the baryon-photon
fluid. Before recombination, successive scatterings destroy the build up of
any polarisation. One therefore anticipates a small degree of polarisation
created {\it at} recombination, partially correlated with the temperature
anisotropies (the velocity part of it). It is convenient to decompose the
polarisation field into two scalar fields denoted E and B (to recall the
similarity of their parity properties with that of the electromagnetic
fields). The power spectrum of the E part is expected to be about 10 times
smaller than for the temperature field T, and the B part {\em which is only
generated by tensor fluctuations} (and by lensing of the CMB by foreground
structures) is yet weaker. The \maps team did not release so far a measurement
of the EE power spectrum, but did provide a measurement of the T-E cross power
spectrum which quantifies the expected correlation of the temperature and E
field.

The WMAP TE spectra turned out to be unexpectedly large at very-low-$\ell$,
which gave strong evidence that the optical depth to the last scattering
surface is rather important ($\sim 0.17$), suggesting that reionisation
happened rather early, around $z\simeq 17$ in typical models. One should note
that this TE spectrum is easier to measure than the EE one owing to the much
larger signal to noise of the temperature, and the cancellation of errors in a
cross-correlations. But if at WMAP sensitivities the TE and EE spectra carry
equal weight (e.g. for constraining the reionisation history), at higher
sensitivity all the information comes from EE. In any case, the important
point to note is that polarisation measurements of the CMB simply probe
various aspects of the same physics that gave rise to the temperature
anisotropies. Such measurements will therefore offer a way to check the
internal consistency of the CMB measurements and help remove degeneracies
present when temperature alone is considered.

More importantly, polarisation measurements will provide the best way to
constrain the scalar to tensor to scalar ratio, $r = A_T/A_S$ (of the
normalisations of the primordial perturbations spectra).  Sufficiently
sensitive measurements will then allow checking the consistency relation
between $r$ and the logarithmic slope of the gravitational wave power
spectrum, $n_T$, which is predicted for inflation. Currently we can only say
that $r < 0.71$ at 95\% confidence limit \cite{0302209}, if only CMB data with
no further priors are used, with essentially no constraint on $n_T$, but the
data will undoubtly continue to improve. Indeed fig.~\ref{fig:sigma_inflat}
shows the current constraints set by WMAP in the $(r, n_s)$ plane, as well as
a forecast for \planck.

So far we concentrated on the case of {\em adiabatic} perturbations which
develop under their own gravity when they are much larger than the
horizon. After they enter the horizon, they become gravitationally stable and
oscillate. All perturbations of the same scale are in phase and started
oscillating (for a given wavelength) at the same time. These leads to the so
called acoustic peaks in the resulting CMB power spectrum. In the case of
active source of fluctuations like in defects theories, perturbations are laid
down at all times, they are generically isocurvature (i.e. it's only the
relative abundance of the fluid components that vary), non-Gaussian, and the
phases of the perturbations of a given scale are incoherent, which does not
lead to many oscillation like in the coherent (\eg inflationary) case. A
single broad peak is expected, and this is why current CMB anisotropy data
already indicates that defects cannot be a major source of CMB fluctuations
and hence cannot seed by themselves the growth of the large scale structures
of the Universe. Still the current CMB data does not yield strong constraints
on which fraction of the perturbations could be isocurvature in nature
(Gaussian or not), while many early universe realistic models naturally lead
to the production of some isocurvature modes. A determination of such
fluctuations, or much tighter upper limit would therefore be extremely
valuable.

Finally one should stress that even if one assumes purely adiabatic initial
conditions, the current data does not allow yet to put strong constraints on
possible deviations from a pure power-law, while measured deviations would
have far-reaching consequences.

One anticipates that the rapid pace of advances from CMB observations will
remain unabated, since the WMAP team should anounce its second data release in
early 2004, it is already known that the BOOMERANG Polarization flight of
January 2003 was a success and the analysis should not be much longer than a
year. In addition, lots of exciting ground and balloon experiments are under
development. Finally, the Planck satellite from ESA is poised for launch in
2007. In order to give an idea of the progresses expected in the next few
years, I now turn to a WMAP/Planck comparison.
 
\section{Planck satellite\label{sec:planck}}

The global similarities and differences of \maps and \plancks are the
following:\begin{itemize}

\item Both map the full sky, from an orbit around the Lagrangian point L2 of
the Sun-Earth system, to minimise parasitic radiation from Earth. Both are
based on the use of off-axis Gregorian telescopes in the 1.5m class. And very
importantly for CMB experiments, both do highly redundant measurements to
better detect and remove (or constrain residuals of) possible systematics
effect, thanks to the long duration of the data taking (at least a year, to be
compared with at most about 10 days for ground experiments which have to cope
in addition with the effect of a changing atmosphere - like ozone clouds, the
closeness to earth, etc\ldots). And both aim at making polarisation
measurements.

\item The American \maps has been designed for rapid implementation, and is
based on fully demonstrated solutions. It's observational strategy uses the
differential scheme. Two telescopes are put back to back and feed differential
radiometers. These radiometers use High Electronic Mobility Transistors
(HEMTs) for direct amplification of the radio-frequency (RF) signal. Angular
resolutions are not better than 10 minutes of arc.

\item The European \plancks is a more ambitious and complex project, which is
to be launched in 2007. It is designed to be the ultimate experiment in
several respects. In particular, several channels of the High Frequency
Instrument (\hfi) will reach the ultimate possible sensitivity per detector,
limited by the photon noise of the CMB itself. Bolometers cooled at 0.1\,K
will allow reaching this sensitivity while, simultaneously, improving the
angular resolution to 5 minutes of arc. The Low Frequency Instrument (\lfi)
limited at frequencies less than 100GHz, will use HEMT amplifiers cooled at 20
Kelvin to increase their sensitivity. The \plancks scan strategy is of the
total power type.  The \lfis uses 4\,K radiative loads for internal references
to obtain this total power measurement. The \hfis readout scheme is based on
an electric modulation of the detector allowing total power measurement.
The combination of these two instruments on \plancks is motivated by the
necessity to map the foregrounds in a very broad frequency range: 30 to 850
GHz.

\end{itemize}
More quantitative aspects are detailed in Table~\ref{tabcaract}, although
these are only indicative since design evolve and in-flight performance may
differ from the requirements (in good or bad, either way are possible).

\begin{table}[ht]
\caption{\label{tabcaract} {\it Planck instrument characteristics.  The
sensitivities ($1\sigma$) are goal values for 12 months integration and for
square pixels whose sides are given in the row "Angular
Resolution". Polarisation measurement at 100 GHz on HFI is waiting for
approval (the sensitivity level without polarisation measurement at 100 GHz is
given in parenthesis). }}  {\small
\begin{center}
    \begin{tabular}[b]{|c||c|c|c||c|c|c|c|c|c|}
	\hline & \multicolumn{3}{c||}{LFI} & \multicolumn{6}{c|}{HFI} \\
	 \hline Detector Technology & \multicolumn{3}{c||}{HEMT arrays} &
	 \multicolumn{6}{c|}{Bolometer arrays} \\ \hline Center Frequency
	 [GHz] & 30 & 44 & 70 & 100 & 143 & 217 & 353 & 545 & 857 \\ Number od
	 Detectors & 4 & 6 & 12 & 8 (4) & 12 & 12 & 6 & 8 & 6 \\ Bandwidth
	 ($\Delta \nu / \nu$) & 0.2 & 0.2 & 0.2 & 0.33 & 0.33 & 0.33 & 0.33 &
	 0.33 & 0.33 \\ Angular Resolution (arcmin) & 33 & 241 & 14 & 9.2 &
	 7.1 & 5.0 & 5.0 & 5.0 & 5.0 \\ $\Delta T/T$ per pixel (Stokes I)
	 [$\mu K/K$] & 2.0 & 2.7 & 4.7 & 2.5 (2.2) & 2.4 & 3.8 & 15 & 17 & 8000
	 \\ $\Delta T/T$ per pixel (Stokes Q and U) [$\mu K/K$] & 2.8 & 3.9 & 6.7
	 & 4.1 (NA) & 4.8 & 7.6 & 30 & \ldots & \ldots \\ \hline
    \end{tabular}
\end{center}
}
\end{table}

\begin{figure}[hbp] \centering \centerline{ \hbox{
\psfig{file=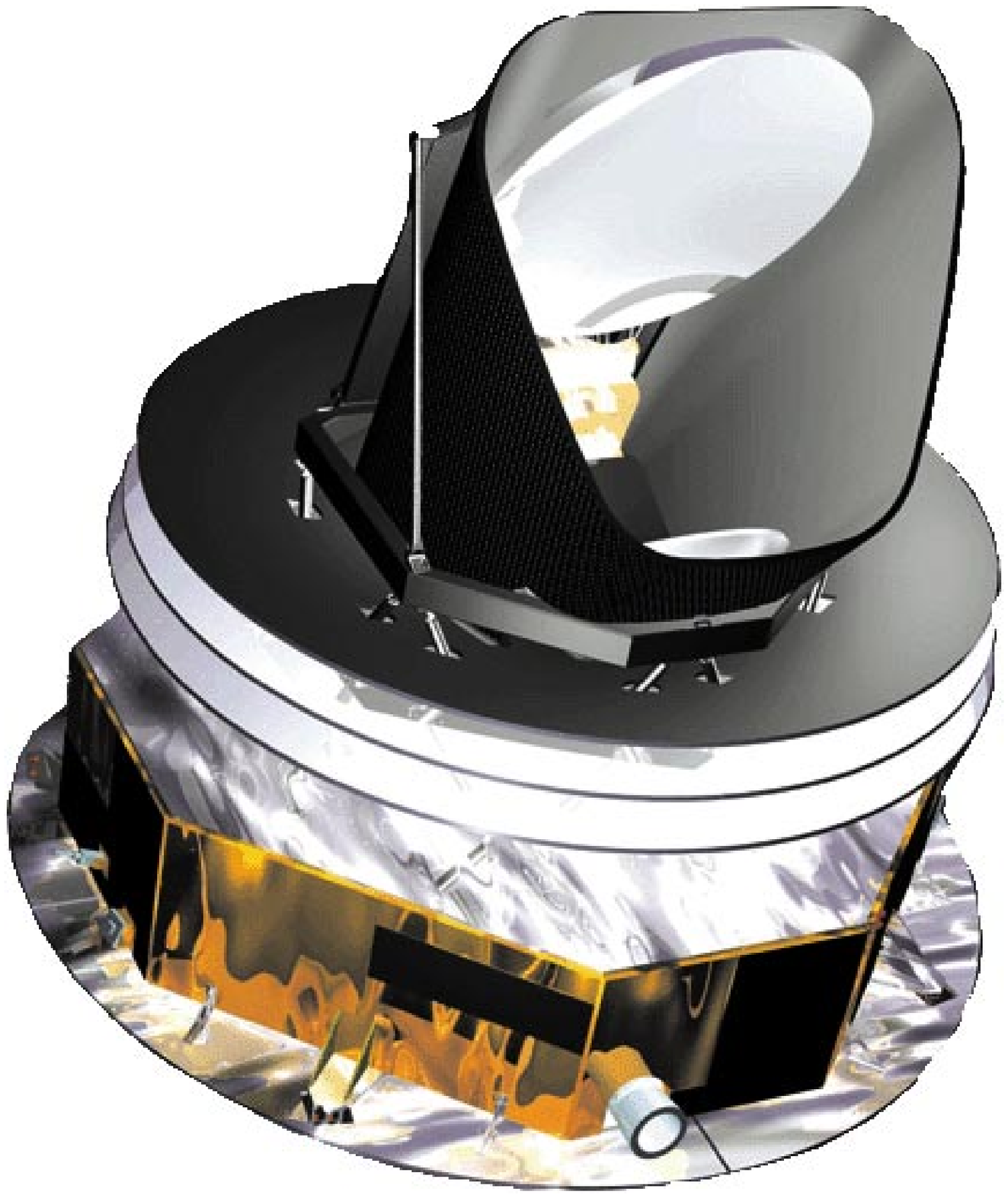, width=0.35\textwidth} \psfig{file=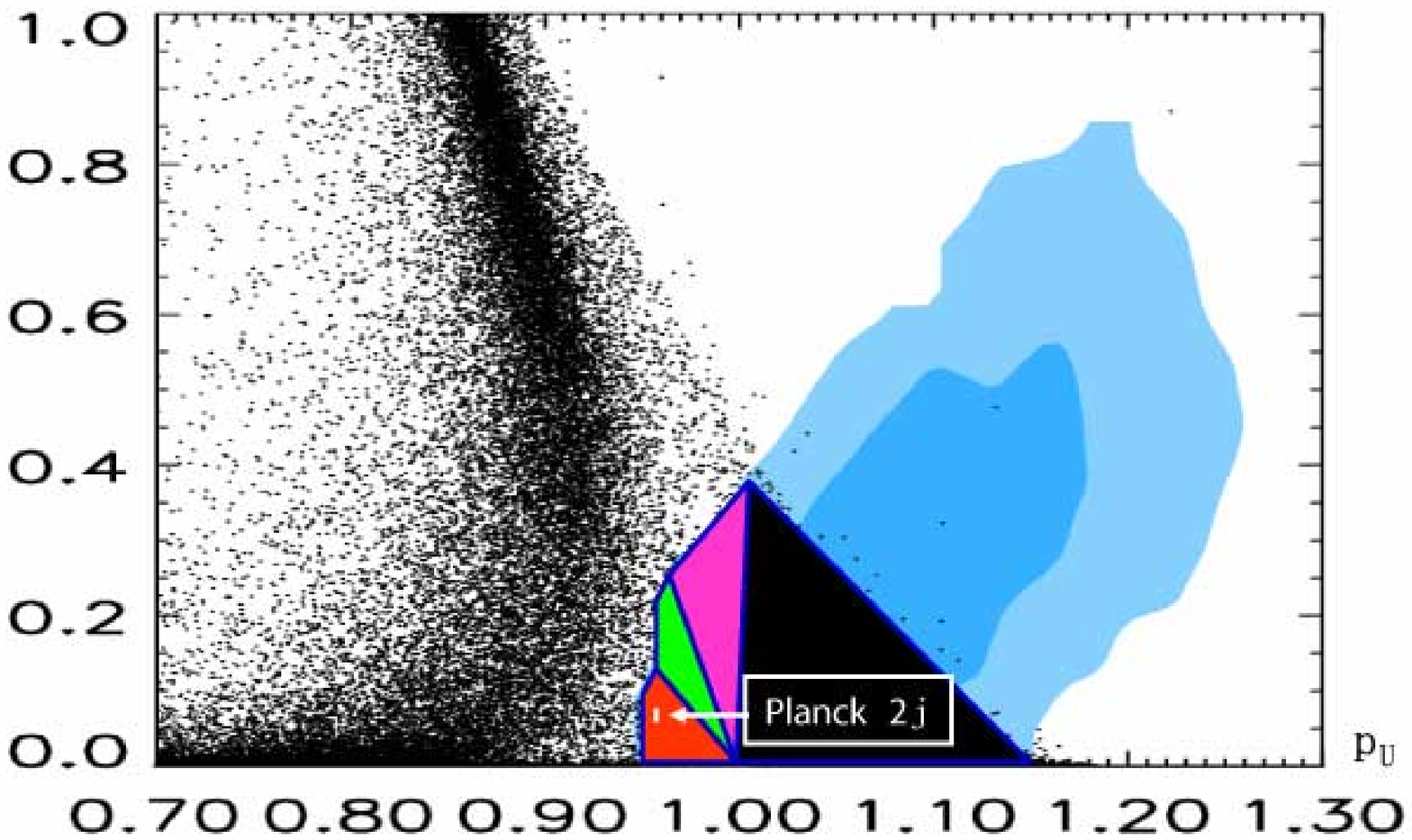,
width=0.65\textwidth, bbllx=95pt, bblly=80pt, bburx=702pt, bbury=476pt}
}}\vspace{-10pt}
\caption[]{a) Artist view of the Planck satellite. b) Constraints in the $(r,
  n_s)$ plane (see text). Each black dot corresponds to 2 parameters of a
  single-field slow-roll inflation models with valid
  dynamics~\cite{2002PhRvD..66h3508K}. The blue shaded regions corresponds to
  the 1 and 2$ \sigma$ constraints from WMAP~\cite{2003ApJS..148..213P}, with
  the red green purple and black overlays each delimiting a class of inflation
  model (see text). The white area illustrate the type of accuracy expected
  from \planck, namely $\Delta r \simeq \Delta n_s \simeq 0.02$.}
\label{fig:sigma_inflat}c
\end{figure}

In order to illustrate how this translates in terms of constraints on early
Universe physics, figure~\ref{fig:sigma_inflat}.b compares the current WMAP
2$\sigma$ constraints (light blue area) versus that anticipated from \planck
(white area), in the plane ($r, n_S$), where $r$ stands for the amplitude of
the primordial tensorial (gravitational wave) power spectrum in units of the
amplitude of the primordial scalar (curvature) power spectrum, and $n_S$
stands for the logarithmic slope of the primordial scalar spectrum (at some
scale). Each black dot corresponds to a couple of reasonable inflation
parameters \cite{2002PhRvD..66h3508K}. In the region of overlap of the dots
cloud with the WMAP-allowed region, the four colour overlays (red, green,
purple and black) each correspond to a particular class of inflation models
ranked by curvature of the potential~\cite{2003ApJS..148..213P}. As
extraordinary as it is to start constraining those elusive but fundamental
parameters, it remains that nearly every inflation model class is still alive
today. But as the white area shows, the Planck data should allow
``zooming-in'' on the parameters of the specific model which will be selected
by the data (if there is such a model, i.e. if the spectra data does not force
us to start considering a broader class than single field slow-roll models).

Before concluding that section, we should recall that the power spectra are
only a first moment (the transform of a 2-pt angular correlation
function). While enough to characterise a fully Gaussian distribution,
deviations from Gaussianity {\em are} expected, albeit at a rather low
level. Such a detection would reveal much about the mechanisms at work in the
early Universe (if they are not residual systematics\ldots).

\section{Conclusions\label{sec:conclu}}

Measurements of the CMB are quite unique in the ensemble of astrophysical
observations that are used to constrain cosmological models. They have the
same character as fundamental physics experiments; they relate fundamental
physical parameters describing our world to well specified signatures which
can be predicted before hand with great accuracy.

The knowledge of CMB anisotropies has literally exploded in the last decade,
since their momentous discovery in 1992 by the DMR experiment on the COBE
satellite. Since then, the global shape of the spectrum has been uncovered
thanks to many ground and balloon experiments and most recently the WMAP
satellite, so far confirming the simplest inflationary model and helping shape
our surprising view of the Universe: spatially flat, and dominated by dark
energy (or $\Lambda$) and cold dark matter, with only a few per cent of
atoms. But the quest is far from over, with many predictions still awaiting to
be checked and many parameters in need of better determination.

If the next 10 years are as fruitful as the past decade, many cosmological
questions should be settled, from a precise determination of all cosmological
parameters to characteristics of the mechanism which seeded the growth of
structures in our Universe, if something even more exciting than what is
currently foreseen does not surge from the future data\ldots

 



\begin{thebibliography}{10}

\bibitem{2000A&A...357....1A}
N.~{Aghanim}, C.~{Balland}, and J.~{Silk}.
\newblock {Sunyaev-Zel'dovich constraints from black hole-seeded
  proto-galaxies}.
\newblock {\em \aap}, 357:1--6, May 2000.

\bibitem{1996A&A...311....1A}
N.~{Aghanim}, F.~X. {Desert}, J.~L. {Puget}, and R.~{Gispert}.
\newblock "{Ionization by early quasars and cosmic microwave background
  anisotropies.}".
\newblock {\em \aap}, 311:1--11, July 1996.

\bibitem{1999A&A...341..640A}
N.~{Aghanim}, F.~X. {Desert}, J.~L. {Puget}, and R.~{Gispert}.
\newblock "erratum: Ionization by early quasars and cosmic microwave background
  anisotropies".
\newblock {\em \aap}, 341:640+, Jan. 1999.

\bibitem{0302207}
C.~L. {Bennett}, M.~{Halpern}, G.~{Hinshaw}, N.~{Jarosik}, A.~{Kogut},
  M.~{Limon}, S.~S. {Meyer}, L.~{Page}, D.~N. {Spergel}, G.~S. {Tucker},
  E.~{Wollack}, E.~L. {Wright}, C.~{Barnes}, M.~R. {Greason}, R.~S. {Hill},
  E.~{Komatsu}, M.~R. {Nolta}, N.~{Odegard}, H.~V. {Peiris}, L.~{Verde}, and
  J.~L. {Weiland}.
\newblock "first year wilkinson microwave anisotropy probe (wmap) observations:
  Preliminary maps and basic results.
\newblock {\em Astrophys. J., in press \& astroph/0302207}, 2003.

\bibitem{0304237}
A.~{Blanchard}, M.~{Douspis}, M.~{Rowan-Robinson}, and S.~{sarkar}.
\newblock {An alternative to the cosmological "concordance model"}.
\newblock {\em astroph/0304237}, 2003.

\bibitem{0305001}
S.~{Boughn} and R.~{Crittenden}.
\newblock {A correlation of the cosmic microwave sky with large scale
  structure}.
\newblock {\em astroph/0305001}, 2003.

\bibitem{1995ApJ...439..503D}
S.~{Dodelson} and J.~M. {Jubas}.
\newblock "reionization and its imprint of the cosmic microwave background".
\newblock {\em \apj}, 439:503--516, Feb. 1995.

\bibitem{0305468}
P.~{Fosalba} and E.~{Gaztanaga}.
\newblock {Measurement of the gravitational potential evolution from the
  cross-correlation between WMAP and the APM Galaxy survey}.
\newblock {\em astroph/0305468}, 2003.

\bibitem{1998ApJ...508..435G}
A.~{Gruzinov} and W.~{Hu}.
\newblock "secondary cosmic microwave background anisotropies in a universe
  reionized in patches".
\newblock {\em \apj}, 508:435--439, Dec. 1998.

\bibitem{2003AnPhy.303..203H}
W.~{Hu}.
\newblock {CMB temperature and polarization anisotropy fundamentals}.
\newblock {\em Annals of Physics}, 303:203--225, Jan. 2003.

\bibitem{1996A&A...315...33H}
W.~{Hu} and M.~{White}.
\newblock "cmb anisotropies in the weak coupling limit.".
\newblock {\em \aap}, 315:33--39, Nov. 1996.

\bibitem{2002PhRvD..66h3508K}
W.~H. {Kinney}.
\newblock {Inflation: Flow, fixed points, and observables to arbitrary order in
  slow roll}.
\newblock {\em \prd}, 66:83508--+, Oct. 2002.

\bibitem{KnoxScoDod.prl}
R.~{Knox}, L.~and{Scoccimarro} and D.~{Dodelson}.
\newblock {\em \prl}, 81, 1998.

\bibitem{0305179}
C.~H. {Lineweaver}.
\newblock "inflation and the cosmic microwave background.
\newblock {\em Proceedings of the New Cosmology Summer School \&
  astroph/0305179}, 2003.

\bibitem{1986ApJ...306L..51O}
J.~P. {Ostriker} and E.~T. {Vishniac}.
\newblock Generation of microwave background fluctuations from nonlinear
  perturbations at the era of galaxy formation.
\newblock {\em \apjl}, 306:L51--L54, July 1986.

\bibitem{PajotThese}
F.~{Pajot}.
\newblock PhD thesis, Paris VII University.

\bibitem{2003ApJS..148..213P}
H.~V. {Peiris}, E.~{Komatsu}, L.~{Verde}, D.~N. {Spergel}, C.~L. {Bennett},
  M.~{Halpern}, G.~{Hinshaw}, N.~{Jarosik}, A.~{Kogut}, M.~{Limon}, S.~S.
  {Meyer}, L.~{Page}, G.~S. {Tucker}, E.~{Wollack}, and E.~L. {Wright}.
\newblock {First-Year Wilkinson Microwave Anisotropy Probe (WMAP) Observations:
  Implications For Inflation}.
\newblock {\em \apjs}, 148:213--231, Sept. 2003.

\bibitem{ReesSciama68}
M.~J. {Rees} and D.~W. {Sciama}.
\newblock Large scale density inhomogeneities in the universe.
\newblock {\em \nat}, 217:511+, 1968.

\bibitem{SachsWolfe67}
R.~K. {Sachs} and A.~M. {Wolfe}.
\newblock {\em \apj}, 147:73, 1967.

\bibitem{1996ApJ...460..549S}
U.~{Seljak}.
\newblock "rees-sciama effect in a cold dark matter universe".
\newblock {\em \apj}, 460:549+, Apr. 1996.

\bibitem{0302209}
D.~N. {Spergel}, L.~{Verde}, H.~V. {Peiris}, E.~{Komatsu}, M.~R. {Nolta}, C.~L.
  {Bennett}, M.~{Halpern}, G.~{Hinshaw}, N.~{Jarosik}, A.~{Kogut}, M.~{Limon},
  S.~S. {Meyer}, L.~{Page}, G.~S. {Tucker}, and E.~{Wollack}.
\newblock "first year wilkinson microwave anisotropy probe (wmap) observations:
  Determination of cosmological parameters".
\newblock {\em Astrophys. J., in press \& astroph/0302209}, 2003.

\bibitem{1980MNRAS.190..413S}
R.~A. {Suniaev} and I.~B. {Zeldovich}.
\newblock "the velocity of clusters of galaxies relative to the microwave
  background - the possibility of its measurement".
\newblock {\em \mnras}, 190:413--420, Feb. 1980.

\bibitem{1972A&A....20..189S}
R.~A. {Sunyaev} and Y.~B. {Zeldovich}.
\newblock "formation of clusters of galaxies; protocluster fragmentation and
  intergalactic gas heating".
\newblock {\em \aap}, 20:189+, Aug. 1972.

\bibitem{1987ApJ...322..597V}
E.~T. {Vishniac}.
\newblock "reionization and small-scale fluctuations in the microwave
  background".
\newblock {\em \apj}, 322:597--604, Nov. 1987.

\bibitem{ZeldovichSunyaev69}
Y.~B. {Zeldovich} and R.~A. {Sunyaev}.
\newblock {\em Astr. Sp. Sci.}, 4:301, 1969.

\end{thebibliography}

\newcommand{\nat}{Nature}

\end{document}